\documentclass[12pt,epsf]{article}
\usepackage{graphicx}
\usepackage{epsfig}
\usepackage{amssymb}
\begin{document}

\def\a{\alpha}
\def\b{\beta}
\def\c{\varepsilon}
\def\d{\delta}
\def\e{\epsilon}
\def\f{\phi}
\def\g{\gamma}
\def\h{\theta}
\def\k{\kappa}
\def\l{\lambda}
\def\m{\mu}
\def\n{\nu}
\def\p{\psi}
\def\q{\partial}
\def\r{\rho}
\def\s{\sigma}
\def\t{\tau}
\def\u{\upsilon}
\def\v{\varphi}
\def\w{\omega}
\def\x{\xi}
\def\y{\eta}
\def\z{\zeta}
\def\D{\Delta}
\def\G{\Gamma}
\def\H{\Theta}
\def\L{\Lambda}
\def\F{\Phi}
\def\P{\Psi}
\def\S{\Sigma}

\def\o{\over}
\def\beq{\begin{eqnarray}}
\def\eeq{\end{eqnarray}}
\newcommand{\gsim}{ \mathop{}_{\textstyle \sim}^{\textstyle >} }
\newcommand{\lsim}{ \mathop{}_{\textstyle \sim}^{\textstyle <} }
\newcommand{\vev}[1]{ \left\langle {#1} \right\rangle }
\newcommand{\bra}[1]{ \langle {#1} | }
\newcommand{\ket}[1]{ | {#1} \rangle }
\newcommand{\EV}{ {\rm eV} }
\newcommand{\KEV}{ {\rm keV} }
\newcommand{\MEV}{ {\rm MeV} }
\newcommand{\GEV}{ {\rm GeV} }
\newcommand{\TEV}{ {\rm TeV} }
\def\diag{\mathop{\rm diag}\nolimits}
\def\Spin{\mathop{\rm Spin}}
\def\SO{\mathop{\rm SO}}
\def\O{\mathop{\rm O}}
\def\SU{\mathop{\rm SU}}
\def\U{\mathop{\rm U}}
\def\Sp{\mathop{\rm Sp}}
\def\SL{\mathop{\rm SL}}
\def\tr{\mathop{\rm tr}}

\newcommand{\bear}{\begin{array}}  
\newcommand {\eear}{\end{array}}
\newcommand{\la}{\left\langle}  
\newcommand{\ra}{\right\rangle}
\newcommand{\non}{\nonumber}  
\newcommand{\ds}{\displaystyle}
\newcommand{\red}{\textcolor{red}}
\def\ubl{U(1)$_{\rm B-L}$}
\def\REF#1{(\ref{#1})}
\def\lrf#1#2{ \left(\frac{#1}{#2}\right)}
\def\lrfp#1#2#3{ \left(\frac{#1}{#2} \right)^{#3}}
\def\OG#1{ {\cal O}(#1){\rm\,GeV}}


\baselineskip 0.7cm

\begin{titlepage}

\begin{flushright}
UT-10-13\\
IPMU 10-0121
\end{flushright}

\vskip 1.35cm
\begin{center}
{\large \bf
Soft Leptogenesis and Gravitino Dark Matter\\ in Gauge Mediation
}
\vskip 1.2cm
Koichi Hamaguchi$^{1,2}$ and Norimi Yokozaki$^1$
\vskip 0.4cm

{\it $^1$ Department of Physics, University of Tokyo,
   Tokyo 113-0033, Japan\\
$^2$ Institute for the Physics and Mathematics of the Universe, 
University of Tokyo,\\ Chiba 277-8568, Japan
}

\vskip 1.5cm

\abstract{
We study soft leptogenesis in gauge mediated supersymmetry breaking models
with an enhanced A-term for the right-handed neutrino.
We find that this scenario can explain the baryon asymmetry of the present universe,
consistently with the gravitino dark matter for a wide 
range of gravitino mass $m_{3/2}={\cal O}({\rm MeV})$--${\cal O}({\rm GeV})$.
We also propose an explicit model which induces the necessary A-term
for the right-handed neutrino.
}
\end{center}
\end{titlepage}

\setcounter{page}{2}

\section{Introduction}\label{sec:1}
Gauge-mediated supersymmetry breaking (GMSB)~\cite{Dine:1981za,Dimopoulos:1981au,Dine:1981gu,Nappi:1982hm,AlvarezGaume:1981wy,Dimopoulos:1982gm,Dine:1993yw,Dine:1994vc,Dine:1995ag,Giudice:1998bp} is an attractive way of communicating supersymmetry (SUSY) breaking effects to the supersymmetric
standard model (SSM), since flavor-changing neutral currents (FCNC) and dangerous CP violating phases are naturally suppressed. 
In addition, the mass spectrum of the superparticles in the SUSY standard model sector is determined by only a few parameters, which may be tested at the LHC in the near future.

In GMSB, the gravitino is the lightest SUSY particle and stable with R-parity. Therefore the gravitino
is a candidate for the dark matter. In fact, gravitinos are produced by the scattering processes of thermal particles after the inflation~\cite{Moroi1993,Bolz2001,Pradler:2006hh}, and its abundance is given by
\begin{eqnarray}
\Omega_{3/2}h^2  &\simeq& 0.4 \times 
\lrfp{m_{3/2}}{0.1\,{\rm GeV}}{-1}
\lrfp{m_{\tilde{g}}}{1\,{\rm TeV}}{2}
\lrf{T_R}{10^7\,{\rm GeV}}\,,
\label{eq:Omega}
\end{eqnarray}
where $\Omega_{3/2}$ and $m_{3/2}$ are the density parameter and the mass of the gravitino, respectively, 
$h\simeq 0.73$ is the normalized Hubble parameter, 
$m_{\tilde{g}}$ is the gluino mass, and $T_R$ is the reheating temperature after the inflation.
Therefore, the gravitino becomes a viable dark matter candidate
 for $T_R\lsim \OG{10^7}\times (m_{3/2}/0.1\,{\rm GeV})$, 
or it can explain the observed cold dark matter density, $\Omega_{\rm CDM}h^2\simeq 0.11$~\cite{Amsler:2008zzb}, if the reheating temperature saturates the bound. 
Furthermore, the notorious inflaton--induced gravitino problem~\cite{Endo:2007sz}, which excludes most of the inflation models in the gravity--mediated SUSY breaking scenario, can be avoided in GMSB models.

However, another big puzzle in cosmology, the origin of the matter anti-matter asymmetry of the universe, is not easy to solve in this framework:
\begin{itemize}
\item
In the standard thermal leptogenesis with heavy right-handed (RH) neutrinos~\cite{Fukugita:1986hr}, there is a lower bound on the mass of the RH neutrino, $M_N \gtrsim 2\times 10^9\,{\rm GeV}$~\cite{Buchmuller:2005eh,Davidson:2008bu}, which requires a high reheating temperature $T_R > \OG{10^9}$. This would lead to a too much gravitino abundance for $m_{3/2}\lsim \OG{10}$ [cf. Eq.~(\ref{eq:Omega})].
\item
Affleck-Dine baryogenesis~\cite{Affleck:1984fy,Dine:1995kz} can work with a low reheating temperature, but in GMSB it generically predicts a stable Q-ball~\cite{Dvali:1997qv,Kusenko:1997si}, and the parameter region is tightly constrained~\cite{Kasuya:2001hg,Takahashi:2010}.
\item
Electroweak baryogenesis~\cite{Cohen:1993nk,Rubakov:1996vz,Cline:2006ts} seems also difficult because 
the necessary ingredients, a first order phase transition and sufficient CP phases, are absent in GMSB.
\end{itemize}

In this paper, we would like to propose a viable baryogenesis scenario in GMSB, 
which is consistent with the gravitino dark matter for a wide range of gravitino mass ${\cal O}({\rm MeV})$--${\cal O}({\rm GeV})$.\footnote{
For other possibilities of baryogenesis and gravitino dark matter in GMSB, see, for instance, 
Refs.~\cite{Asaka:1999jb,Kasuya:2001hg,Hamaguchi:2001gw,Fujii:2002yx,Fujii:2002fv,Fujii:2003iw,Lemoine:2005hu,Olechowski:2009bd,Shoemaker:2009kg}.}
The framework is a simple GMSB model supplemented by an enhanced A-term for the RH neutrino,
and the baryon asymmetry is produced by the soft leptogenesis~\cite{Grossman2003,D'Ambrosio2003}.

Soft leptogenesis~\cite{Grossman2003,D'Ambrosio2003} is an attractive way of generating baryon asymmetry. The SUSY breaking terms introduce  a
mixing between the RH sneutrinos and their anti-particles. 
This induces significant CP violation in sneutrino decays in similar ways to $B^0$-$\bar{B}^0$ and $K^0$-$\bar{K}^0$ mixings.
An attractive feature of the soft leptogenesis is that $M_N$ (and $T_R$)
can be smaller than that in the standard leptogenesis and 
therefore there is a possibility of generating baryon asymmetry without 
generating too much gravitino dark matter. 

Interestingly, a successful soft leptogenesis favors
a small B-term for the RH neutrino,
which is naturally realized in the framework of GMSB~\cite{Grossman2005}.
In Ref.~\cite{Grossman2005}, 
the authors investigated the soft leptogenesis in a minimal GMSB setup,
and found a viable parameter region with very light gravitino $m_{3/2} \lesssim 16\, {\rm eV}$. 
In the minimal setup, the RH neutrino A-term is suppressed, 
and hence sufficient baryon asymmetry cannot be generated for $m_{3/2}\gtrsim {\cal O}(100\,{\rm eV})$
satisfying the gravitino constraint.
We extend their study with an enhanced A-term,
and show that there is a viable region with $m_{3/2}={\cal O}({\rm MeV})$--${\cal O}({\rm GeV})$,
consistently with the gravitino dark matter.
We also show  an explicit model which generates an enhanced A-term
through the coupling between the messenger and up-type Higgs, without introducing additional unwanted CP phases in the low energy.

This paper is organized as follows: 
In section \ref{sec:soft}, we briefly review soft leptogenesis,
and then show that
the baryon asymmetry can be explained in our scenario.
In section \ref{sec:model}, 
we introduce a concrete model which generates the necessary A-term 
through a coupling between the up-type Higgs and the messenger. 
Section \ref{sec:sum} is devoted to summary and discussion.

\section{Soft Leptogenesis}\label{sec:soft}
Let us first briefly review the soft leptogenesis following Ref. \cite{D'Ambrosio2003}. 
We consider only the lightest RH neutrino and sneutrino for simplicity. 
The superpotential for the RH neutrino is given by
\begin{eqnarray}
 W = \frac{1}{2} M_N N N + Y_{\nu,i} L_i H_u N ,
\end{eqnarray}
where $N$, $L_i$ and $H_u$ are the chiral superfields for the RH neutrino, the lepton doublets, the up-type Higgs, respectively. 
The soft SUSY breaking terms containing RH sneutrino $\tilde{N}$ are
\begin{eqnarray}
 -\mathcal{L}_{soft} = m_{\tilde{N}}^2\tilde{N}^{*} \tilde{N} + \frac{B_{\nu} M_N}{2} \tilde{N} \tilde{N} + A_{\nu} Y_{\nu,i} \tilde{L}_i H_2 \tilde{N} + h.c. .
\end{eqnarray}
$M_N$ and $Y_{\nu,i}$ are taken to be real by redefining the phases of the superfields, $N$ and $L_i$. 
SUSY breaking terms introduce the mixing between $\tilde{N}$ and $\tilde{N}^*$, 
which induces lepton asymmetry in decays of $\tilde{N}$ and $\tilde{N}^*$ .
Then,
the generated lepton asymmetry is converted into the baron asymmetry through the sphaleron process~\cite{Kuzmin:1985mm}.
The baryon to entropy ratio is given by \cite{D'Ambrosio2003}\footnote{%
Here and hereafter, we neglect flavor effects \cite{Fong:2008mu}, quantum effects \cite{Fong:2008yv},
and corrections suppressed by ${\cal O}(m_{\rm soft}/M_N)^2$~\cite{Grossman2004}, 
for simplicity. These effects are small in most of the parameter region of our interest.}
\begin{eqnarray}
 \frac{n_B}{s} 
&\simeq& 8.6 \times 10^{-4} \left[\frac{4\Gamma |B_\nu|}{4|B_\nu|^2 + \Gamma^2}\frac{|A_\nu|\sin\theta}{M_N} \right] \eta 
, \label{eq:nb}
\end{eqnarray}
where $\Gamma$ is the width of the RH sneutrinos
\begin{eqnarray}
 \Gamma \simeq \frac{|Y_{\nu,i}|^2}{4\pi} M_N\,,
\end{eqnarray}
$\theta$ is the CP phase given by
 $\theta={\rm arg}(B_\nu A_\nu^*)$, 
 and $\eta$ is a factor which describes the effects caused by the inefficiency in the production of the RH sneutrinos,
 the wash-out effects, and the temperature dependence of the
 phase space of fermionic and bosonic final states in RH sneutrino decays.
The maximum value of $\eta$ is $\mathcal{O}(0.1)$ for $\widetilde{m_1} \simeq 10^{-(3-4)}{\rm eV}$~\cite{D'Ambrosio2003}, where
\begin{eqnarray}
\widetilde{m_1} = \frac{|Y_{\nu,i}|^2 v^2}{M_N},
\end{eqnarray}
and $v=174\,{\rm GeV}$ is the vacuum expectation value of Higgs.

\begin{figure}[t!]
\begin{center}
\includegraphics[width=10cm]{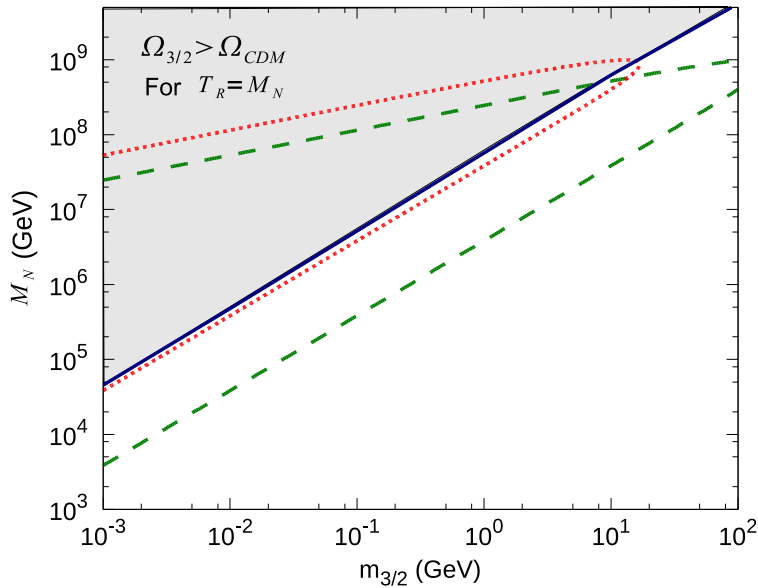}
\end{center}
\caption{The region with successful baryogenesis is shown in $m_{3/2}-M_M$ plane. Sufficient baryon asymmetry
 can be generated in the region enclosed by the red dotted line (green dashed line) for $B_\nu = 0.1 m_{3/2} \, (B_\nu =0.01 m_{3/2})$.
The gray region above the blue solid line is excluded due to too large abundance of the gravitino for $T_R = M_N$.
 We take $A_\nu=1\,{\rm TeV}$ and $m_{\tilde{g}} = 750\, \GEV$.
}
\label{fig:nbs}
\end{figure}

Now let us estimate the baryon asymmetry by using Eq.~(\ref{eq:nb}) in our setup. 
For simplicity, we assume $\widetilde{m_1} \simeq 10^{-3} {\rm eV}$ and $\eta\simeq 0.1$,\footnote{
The precise value of $\eta$ depends on the initial abundance of the RH sneutrino~\cite{D'Ambrosio2003}.
} which leads to
\begin{eqnarray}
\Gamma \;\simeq\; \frac{\widetilde{m_1}}{4\pi v^2}M_N^2
 \;\simeq\; 0.26~{\rm MeV}\lrfp{M_N}{10^7\,{\rm GeV}}{2}\,.
 \label{eq:GMN}
\end{eqnarray}
We also assume that $B_\nu$ is dominated by the gravity-mediation contribution, $|B_\nu|\sim m_{3/2}$,\footnote{
The contribution from the anomaly mediation is also included.}
 as discussed in Ref.~\cite{Grossman2005}. Fig.~\ref{fig:nbs} shows the region where a sufficient baryon asymmetry is generated in $m_{3/2}$--$M_N$ plane.
In the region enclosed by the red dotted line (green dashed line), $n_B/s$ can explain the observed value 
$[n_B/s]_{\rm obs} \simeq 8.7 \times 10^{-11} $ \cite{Amsler:2008zzb} for
$B_\nu = 0.1 m_{3/2} \, (B_\nu =0.01 m_{3/2})$.
Here, we have taken $A_\nu \simeq  1~{\rm TeV}$ (see discussion below).
The upper (red dotted and greed dashed) lines show the upper bounds on $M_N$,
$M_N^{\rm max}\propto B_\nu^{1/3}$, and the lower lines show the lower bounds
$M_N^{\rm min}\propto B_\nu$. 
These behaviors can be understood 
from the fact that $n_B/s \propto M_N^{-3} B_\nu$ for $|B_\nu| \ll \Gamma$
and $n_B/s \propto M_N B_\nu^{-1}$ for $|B_\nu| \gg \Gamma$ [cf. Eqs.~(\ref{eq:nb})(\ref{eq:GMN})].

A successful leptogenesis requires $T_R\gsim M_N$. 
On the other hand, the maximal value of $T_R$, which is consistent with the dark matter abundance, is given by the 
requirement that $\Omega_{3/2} < \Omega_{CDM}$ [cf. Eq.(\ref{eq:Omega})]. This leads to 
\begin{eqnarray}
 M_N \lesssim \mathcal{O}(10^7 \, \GEV) \left(\frac{m_{3/2}}{0.1\, \GEV}\right) \label{eq:mnmax}.
\end{eqnarray}
In Fig.~\ref{fig:nbs}, we have also shown the constraint $\Omega_{3/2} h^2 < 0.121$~\cite{Amsler:2008zzb} as 
a blue solid line, for $T_{R} = M_N$ and $m_{\tilde{g}}=750$ GeV.\footnote{%
We have included Bino and Wino contributions to the gravitino production~\cite{Pradler:2006hh},
assuming the GUT relation among gaugino masses.}
Around the blue solid line, the gravitino can be the dominant component of the dark matter. 

Now we estimate the required size of $A_\nu$. As discussed above, smaller $M_N$ 
is favored by the constraint from 
the gravitino abundance,
which corresponds to $\Gamma < |B_\nu|$ [cf. Eq.(\ref{eq:GMN})].
By taking $\Gamma\ll |B_\nu|$, we obtain
\begin{eqnarray}
 \frac{n_B}{s} \simeq 2.2 \times 10^{-19} \, \frac{M_N}{|B_\nu|} \left(\frac{|A_\nu|}{1 \TEV}\right) .
\end{eqnarray}
{}From Eq. (\ref{eq:mnmax}), this implies
\begin{eqnarray}
 {\rm max}\left(\frac{n_B}{s}\right) \simeq {\cal O}(10^{-11}) \left(\frac{m_{3/2}}{|B_\nu|}\right) \left(\frac{|A_{\nu}|}{1\TEV}\right) .
\end{eqnarray}
Therefore in order to explain the observed value of $n_B/s$,
 $|A_\nu|\simeq (100\, \GEV-10\, \TEV)$ is required for $|B_\nu| \simeq (0.01-1) m_{3/2}$.
However, such a large A-term is not generated in a minimal GMSB. In fact, it was found~\cite{Grossman2005} that
a successful soft leptogenesis and the gravitino constraint require an ultralight gravitino $m_{3/2}\lsim 16$~eV, as far as the A-term is generated through the renormalization--group evolutions. In the next section, we show an explicit model which can generate a large A-term for the RH neutrino.

\section{A Model}\label{sec:model}
In this section, we give a concrete model which generates the enhanced $A_\nu$, through
a new coupling between the messenger field and the up-type Higgs. We  will also
show that this coupling does not induce large CP violation.

In GMSB, the messenger mass is given by the following superpotential:
\begin{eqnarray}
 W = X  \Psi \bar{\Psi},
\end{eqnarray}
where $\Psi$ and $\bar{\Psi}$ are messenger superfields and transform as ${\bf 5}$ and ${\bf \bar{5}}$ under the GUT SU$(5)$, respectively. $X$ is 
a superfield which has a scalar and an auxiliary vacuum expectation values,
$\left<X\right>=M+F\theta^2$. 
 We consider only one pair of messengers for simplicity, 
however an extension  to the multi-messenger case is straightforward.

\begin{figure}[t!]
\begin{center}
\includegraphics[width=7cm]{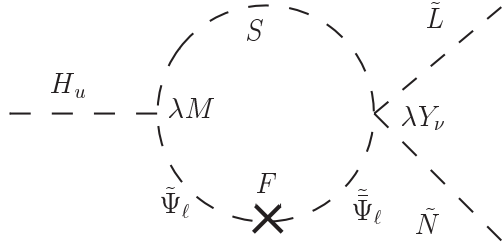}
\end{center}
\caption{The Feynman diagram generating $A_\nu$.}
\label{fig:a-term}
\end{figure}

In order to generate the A-term, $\mathcal{L} = - A_{\nu} Y_{\nu} \tilde{L} H_u \tilde{N} + h.c.$,
 we consider the following superpotential \cite{Giudice1998,Endo2007}:
\begin{eqnarray}
 W = \lambda S H_u \bar{\Psi}_{\ell} \,, \label{eq:model}
\end{eqnarray}
where $\bar{\Psi}_{\ell}$ is the leptonic part of the messenger $\bar{\Psi}$ and $S$ is a gauge singlet. In order to forbid
 unwanted terms, such as $N H_u \bar{\Psi}_{\ell}$ and $N H_d {\Psi}_{\ell}$,\footnote{If these term exist, the large $B_\nu$ can be generated.}
 we introduce 
a messenger parity $Z'_2$ in addition to the R-parity (equivalently matter parity). 
$S, \Psi$ and $\bar{\Psi}$ are odd and the others are even under $Z'_2$.
There is also another term, $\lambda' S H_d \Psi_{\ell}$, which is allowed by the symmetry.
However, this term is irrelevant to the generation of $A_\nu$,
and therefore we neglect it in following discussion, for simplicity.

We take $\lambda$ and the messenger mass $M$ to be real by redefining the phases of $\Psi$ and $\bar{\Psi}$. 
We assume that the mass of $S$, $M_S$ satisfy the relation, $M_N < M_S < M$. 
The term in Eq.({\ref{eq:model}}) decouples after integrating out the messenger superfields.
The leading contribution to the A-term which is proportional to $(F/M)$ is given by the one-loop diagram expressed in Fig. \ref{fig:a-term}.
This contribution is also extracted from the wave-function renormalization of $H_u$ by the analytic continuation method \cite{Giudice1998}. 
The leading term of $A_\nu$ is given by
\begin{eqnarray}
A_{\nu} \simeq \left. \frac{\partial \ln Z_{H_u}}{\partial \ln X} \right|_{X=M} \frac{F}{M} = -\frac{\lambda^2}{16\pi^2}\frac{F}{M} , \label{eq:a_nu}
\end{eqnarray}
which can be of the order of $100\, \GEV-1\, \TEV$.

Next we discuss the possible source of CP violation induced by the new coupling in Eq. (\ref{eq:model}). 
In addition to $A_\nu$, this coupling generates A-terms for up-type squarks $A_u$, Higgs B-term $B$ 
and soft SUSY breaking mass for up-type Higgs $m_{H_u}^2$ and squarks $m_{\tilde{Q}}^2$, $m_{\tilde{\bar{U}}}^2$. 
The $m_{H_u}^2$ , $m_{\tilde{Q}}^2$
and $m_{\tilde{\bar{U}}}^2$ do not induce an additional CP phase
beyond the CKM phase. $A_u$ and $B$ are given by
\begin{eqnarray}
 A_u = B \simeq -\frac{\lambda^2}{16\pi^2}\frac{F}{M}. \label{eq:b_mu}
\end{eqnarray}
The corresponding terms in the superpotential and soft breaking terms are defined by
\begin{eqnarray}
 W = Y_{u,ij} Q_i H_u \bar{U}_j - \mu H_d H_u, \ \ 
 -\mathcal{L}_{soft} = A_u Y_{u,ij} \tilde{Q}_i H_u \tilde{\bar{U}}_j - B \mu H_d H_u + h.c. .
\end{eqnarray}
We assume that the Higgs $\mu$ term is generated above the messenger scale. Under this assumption, there is no physical phase from 
GMSB, since the phases of the soft breaking parameters are the same, ${\rm arg(F/M)}$ and we can remove them by the $U(1)_R$ transformation. 
On the other hand, the neutrino B-term, $B_\nu$ is generated by the gravity-mediation and the order of gravitino mass. Therefore its phase is
 expected to be completely different from ${\rm arg}(F/M)$. With the $U(1)_R$ transformation and a phase transformation of $H_u$, the parameters transform as,
\begin{eqnarray}
&& \mu \rightarrow \mu e^{i(\theta_{H_u} -2 \theta_R) }, \ \ B_\nu \rightarrow B_{\nu} e^{2i \theta_R}, \nonumber \\
&& A_{\nu,u} \rightarrow A_{\nu,u} e^{2i \theta_R}, \ \  B\mu \rightarrow B \mu e^{i \theta_{H_u}} .
\end{eqnarray}
If we choose $2\theta_R = - {\rm arg}(F/M)$ and $\theta_{H_u} = -{\rm arg}(\mu)-{\rm arg}(F/M)$, only $B_\nu$ is complex and its phase is
${\rm arg}(B_\nu F^*/M)$. Therefore the new interaction term does not lead to large CP violation in low energy phenomena.

\section{Summary and Discussion}\label{sec:sum}

We considered soft leptogenesis in gauge mediated SUSY breaking scenario, including the simple interaction term
 which contains up-type Higgs and leptonic part of the messenger. The interaction term generates $A_\nu$, $A_u$,
 soft SUSY breaking mass for up-type Higgs and squarks, and Higgs B-term.
With the large $A_\nu$ soft leptogenesis works successfully, which is consistent with the
gravitino dark matter for a wide range of gravitino mass.
The phases of $A_\nu$, $A_u$ and Higgs B-term are aligned with those of other SUSY breaking terms from gauge mediated
 SUSY breaking. Therefore inclusion of the new interaction term does not lead to large CP violation in low energy phenomena.
  Interestingly, the additional contributions to the soft terms
 lead to a different spectrum pattern of SUSY particles from that of ordinary gauge mediation,
 which may be tested at LHC in the near future.

\section*{Acknowledgements}
We would like to thank M. Endo, T. Moroi and F. Takahashi for useful discussions.
NY is supported by Grand-in-Aid for Scientific Research, No.22-7585 from JSPS, Japan.
The work of K.H. was supported by JSPS Grant-in-Aid for Young Scientists (B) (21740164) and
Grant-in-Aid for Scientific Research (A) (22244021).
This work was supported by World Premier International Center Initiative (WPI Program), MEXT, Japan.

\providecommand{\href}[2]{#2}\begingroup\raggedright\endgroup

\end{document}